\documentclass[11pt]{article}

\usepackage[T1]{fontenc}
\usepackage[utf8]{inputenc}
\usepackage{lmodern}
\usepackage[a4paper,margin=1in]{geometry}
\usepackage{amsmath,amssymb,amsfonts}
\usepackage{braket}
\usepackage{graphicx}
\usepackage{booktabs}
\usepackage{tabularx}
\usepackage{array}
\usepackage{makecell}
\usepackage{float}
\usepackage{enumitem}
\usepackage{cite}
\usepackage{url}
\usepackage{tikz}
\usetikzlibrary{arrows.meta,positioning,shapes.geometric,decorations.pathmorphing,calc}
\usepackage[hidelinks]{hyperref}
\newcolumntype{P}[1]{>{\raggedright\arraybackslash}p{#1}}

\title{A Bell-State Extension of Loop-Back Quantum Key Distribution}
\author{%
Luis Adri\'an Lizama-P\'erez\\[4pt]
\small Departamento de Sistemas de Informaci\'on y Comunicaciones,\\
\small Divisi\'on de Ciencias B\'asicas e Ingenier\'ia,\\
\small Universidad Aut\'onoma Metropolitana, Unidad Lerma,\\
\small Av. de las Garzas No.~10, Col. El Pante\'on,\\
\small Lerma, Estado de M\'exico 52005, Mexico\\
\small \texttt{l.lizama@correo.ler.uam.mx}\\[4pt]
\small\emph{Previous affiliation:}\\
\small Department of Electronics, Universidad T\'ecnica Federico Santa Mar\'ia,\\
\small Santiago, Chile.%
}
\date{}

\begin{document}
\maketitle

\begin{abstract}
Bidirectional quantum key distribution (QKD) protocols face persistent challenges related to classical disclosure, confinement of the signal space to predictable subspaces, and limited detectability under substitution or entanglement--swapping attacks.
In this work, we present a Bell--state extension of the loop--back QKD architecture that improves efficiency and detectability while preserving its defining feature of a simplified, measurement--free remote terminal.
The protocol employs entangled Bell states together with deterministic local Pauli encoding at the remote node.
A central element is that Alice privately prepares and knows the initial Bell state, which serves as a hidden reference enabling her to interpret the Bell--state transition induced by Bob, while preventing an adversary from reconstructing the encoding without access to this reference.
By exploiting both intra-- and inter--family Bell transitions, the scheme expands the effective signal space beyond the subspace restrictions of earlier two--way protocols.
Alice performs a Bell--state measurement to deterministically infer Bob's operation without any basis sifting.
Although the traveling subsystem remains locally maximally mixed, concealing the initial Bell family amplifies disturbance under separable substitution strategies, yielding an intrinsic detection probability of approximately $3/4$ per round.
From an efficiency perspective, the protocol lifts the intrinsic post--selection limitation of single--qubit loop--back schemes: the effective throughput is bounded only by the Bell--state measurement success probability, reaching up to $50\%$ in linear--optical implementations.
These features make the proposed scheme particularly suitable for mobile or edge--based QKD scenarios, where passive remote nodes must operate under high loss and limited interaction times.
\end{abstract}


\section{Introduction}

Quantum key distribution (QKD) has become a central tool for secure communication, supported by extensive theoretical and experimental advances~\cite{pirandola2020advances}. 
Beyond one--way paradigms such as BB84, bidirectional QKD architectures have attracted sustained interest due to their potential for deterministic inference, intrinsic channel verification, and asymmetric hardware deployment. 
However, two--way protocols continue to face persistent challenges, including:
(i) significant classical disclosure during basis reconciliation,
(ii) confinement of information to predictable subspaces exploitable by two--way or side--channel attacks~\cite{braunstein2012side}, and
(iii) limited detectability under substitution or entanglement--swapping strategies~\cite{pan1998experimental,zukowski1993event}.
Addressing these limitations is essential for developing deterministic and low--overhead quantum communication models compatible with modern integrated and mobile photonic platforms~\cite{wang2020integrated}.

Several deterministic frameworks have been proposed to mitigate these issues.
The Ping--Pong protocol~\cite{bostrom2002deterministic} pioneered entanglement--based two--way communication by encoding information through intra--family transformations of a publicly known Bell state.
This reliance on a fixed and disclosed Bell family enables coherent attacks that preserve the subspace structure~\cite{wojcik2003eavesdropping,cai2006eavesdropping,zhang2004improving}, a vulnerability later confirmed in comprehensive analyses of two--way QKD security~\cite{beaudry2013security}.
Lucamarini and Mancini subsequently introduced the LM05 protocol~\cite{lucamarini2005secure}, which removes the need for entanglement but requires active unitary modulation at the remote node and remains vulnerable to Trojan--horse and loss--based attacks unless strong optical isolation is enforced~\cite{gisin2006trojan}.

More recently, the Loop--Back QKD protocol~\cite{lizama2025loop} introduced a bidirectional architecture in which a single quantum pulse travels forth and back through the same channel, while all quantum detection and post--processing are concentrated at Alice's station.
Bob acts as a simplified and measurement--free terminal, implementing only a passive polarization operation and reflection.
This asymmetric design eliminates detectors at the network edge, reduces synchronization requirements, and enables compact, low--power implementations suitable for quantum--mobile and IoT scenarios.
An extended three--basis configuration was subsequently reported~\cite{lizama2025three}, preserving the passive nature of Bob while improving noise tolerance and statistical robustness.
Despite these advantages, existing Loop--Back realizations rely on single--qubit encodings, which restrict the accessible Hilbert space and yield an intrinsic post--selection efficiency of approximately $25$--$26\%$, while its single-qubit encoding space motivates extensions that leverage entanglement correlations to further expand the effective signal space and strengthen robustness against broader substitution strategies. In mobile or dynamically reconfigurable scenarios, where channel losses and temporal availability are inherently more severe, this intrinsic efficiency ceiling becomes a limiting factor for practical key generation.

While the resulting asymmetric architecture shares the objective of detector simplification with measurement-device-independent QKD~\cite{lo2012measurement}, the present protocol differs fundamentally in that it relies on bidirectional propagation and deterministic Bell-state inference rather than relay-based measurement.

In this work, we introduce a Bell--state extension of the Loop--Back QKD architecture.
The protocol preserves the defining features of Loop--Back schemes---namely bidirectional propagation and a simplified, measurement--free remote terminal---while extending the architecture into the entangled--state regime.
A central element of the proposal is that Alice privately prepares and retains knowledge of the initial Bell state, which acts as a hidden reference for interpreting the evolution of the shared state after Bob's operation.

\subsection*{Contributions of This Work}

The present work makes the following contributions to the development of bidirectional and passive QKD architectures:

\begin{enumerate}
    \item We extend the Loop--Back QKD architecture to Bell states while preserving its asymmetric and detector--free remote terminal, thereby expanding the accessible signal space without increasing hardware complexity at the network edge.

    \item We introduce a concealed Bell--state reference, known only to Alice, which removes any public frame for interpreting Bell--state transitions and prevents family--preserving or symmetry--exploiting strategies affecting earlier two--way and Loop--Back protocols.

    \item By enabling both intra--family $(I,Z)$ and inter--family $(X,Y)$ transitions within the Bell basis, the protocol lifts the intrinsic efficiency limitation of single--qubit Loop--Back schemes. The effective throughput is bounded by $p_{\mathrm{BSM}}$. Standard linear--optical implementations reach $p_{\mathrm{BSM}}\leq 50\%$~\cite{calsamiglia2001maximum,bayerbach2023bell}.

    \item We show that the resulting four--dimensional Bell space enhances the detectability of separable substitution strategies, yielding an intrinsic disturbance sensitivity on the order of $3/4$ per round while remaining compatible with current and near--term experimental platforms.
\end{enumerate}

Operationally, Bob applies a local Pauli operation that induces either intra--family or inter--family transitions within the Bell basis and returns the traveling subsystem to Alice.
A Bell--state measurement (BSM) performed at Alice's station then enables deterministic identification of Bob's operation without any basis--sifting stage, leveraging well--established Bell--state discrimination techniques~\cite{pan2012multiphoton} and the underlying dense-coding correlation structure of Bell states~\cite{bennett1992communication}.
Throughout the protocol, the traveling subsystem remains locally maximally mixed, as in standard entanglement--based schemes.

As a consequence of this design, the protocol eliminates basis reconciliation, avoids confinement to publicly known subspaces, and exhibits enhanced sensitivity to substitution strategies.
In contrast to single--qubit Loop--Back realizations, the effective efficiency is no longer limited by post--selection in a two--dimensional space, but instead determined by the Bell--state measurement success probability $p_{\mathrm{BSM}}$, which reaches up to $50\%$ in linear--optical implementations~\cite{bayerbach2023bell}.

To the best of our knowledge, no existing Loop--Back protocol exploits inter--family Bell transitions or conceals the reference Bell state.
By extending the Loop--Back architecture into the entangled--state regime, the present work establishes a direct architectural bridge between passive two--way QKD and Bell--state--based inference, offering a pathway to improved efficiency and robustness while remaining compatible with current photonic technology.

The remainder of this paper is organized as follows.
Section~2 describes the proposed Bell--state Loop--Back protocol and its operational structure.
Section~3 analyzes the algebraic framework and security properties of the scheme.
Section~4 outlines future work and perspectives for experimental and theoretical extensions.
Section~5 concludes the paper with a summary of the main contributions and final remarks.

\section{Loop-Back Bell-State Protocol}

This section presents the proposed \textit{Loop--Back Bell-State QKD} protocol, describing its operational sequence, unitary encoding mechanism, and verification procedure within a bidirectional architecture with a measurement--free remote terminal.

\subsection*{Protocol Overview}

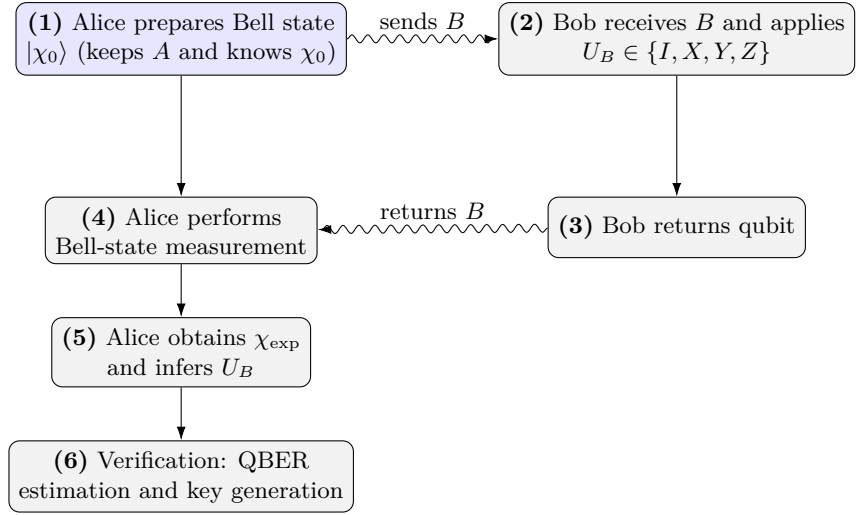
\begin{figure}[h!]
\centering
\begin{tikzpicture}[
    node distance=0.9cm and 1.0cm,
    every node/.style={font=\footnotesize, align=center},
    block/.style={rectangle, rounded corners, draw, minimum width=2.8cm, minimum height=0.8cm, fill=gray!10},
    line/.style={draw, -{Latex[length=1.8mm,width=1.2mm]}},
    qline/.style={draw, -{Latex[length=1.8mm,width=1.2mm]}, decorate,
                   decoration={snake, amplitude=1.5pt, segment length=6pt}}
]

\node[block, fill=blue!10] (aliceprep) 
{\textbf{(1)} Alice prepares Bell state \\ $\ket{\chi_0}$ (keeps $A$ and knows $\chi_0$)};
 
\node[block, right=2.0cm of aliceprep] (bobop) 
{\textbf{(2)} Bob receives $B$ and applies \\ $U_B\in\{I,X,Y,Z\}$};

\node[block, below=1.6cm of bobop] (bobret) 
{\textbf{(3)} Bob returns qubit};

\node[block, below=1.6cm of aliceprep] (alicemeas) 
{\textbf{(4)} Alice performs \\ Bell-state measurement};

\node[block, below=0.7cm of alicemeas] (aliceinf) 
{\textbf{(5)} Alice obtains $\chi_{\mathrm{exp}}$ \\ and infers $U_B$};

\node[block, below=0.7cm of aliceinf] (verify) 
{\textbf{(6)} Verification: QBER \\ estimation and key generation};

\path[qline] (aliceprep) -- (bobop)
    node[midway, above]{sends $B$};

\path[line] (bobop) -- (bobret);

\path[qline] (bobret) -- (alicemeas)
    node[midway, above]{returns $B$};

\path[line] (aliceprep) -- (alicemeas);
\path[line] (alicemeas) -- (aliceinf);
\path[line] (aliceinf) -- (verify);

\node[above=0.15cm of aliceprep, font=\small\bfseries] 
{Loop-Back Bell-State QKD Protocol (Full-Pauli Encoding)};

\end{tikzpicture}

\caption{Compact flow diagram of the \textit{Loop-Back Bell-State QKD} protocol with full-Pauli encoding.
The numbered blocks correspond to Steps~(1)--(6) described in Section~2.
Alice privately prepares the initial Bell state $\ket{\chi_0}$ and sends one qubit to Bob.
Bob applies a local Pauli operation and returns the qubit through the same quantum link.
Alice performs a Bell-state measurement, deterministically infers the applied operation, and uses a subset of rounds for verification and key generation.
The fact that $\chi_0$ is chosen and known exclusively by Alice reinforces the indistinguishability of the induced transition for any external observer.}
\label{fig:bell_loopback_flow}
\end{figure}

For clarity, the protocol is described as a sequence of operational steps repeated over many independent rounds. 
Each step is illustrated schematically in Fig.~\ref{fig:bell_loopback_flow} and detailed below.

\subsection*{Step 1: Preparation of the Initial State}

A quantum source prepares a pair of qubits in a Bell state chosen privately by Alice,
\[
\ket{\chi_0}\in 
\bigl\{\ket{\Phi^{+}},\ket{\Phi^{-}},\ket{\Psi^{+}},\ket{\Psi^{-}}\bigr\},
\]
and distributes the qubits between Alice ($A$) and Bob ($B$).
Alice retains qubit $A$ and sends qubit $B$ to Bob through a quantum channel.
The private selection of $\ket{\chi_0}$ conceals the Bell-family reference and prevents any external party from exploiting intra-family symmetries.
Any manipulation of the distributed state, including untrusted-source or substitution scenarios, alters the expected Bell correlations and becomes detectable once Alice performs a Bell-state measurement.

\subsection*{Step 2: Bob's Unitary Encoding}

Upon receiving qubit $B$, Bob randomly selects a Pauli operator
\[
U_B \in \{I, X, Y, Z\},
\]
and applies it locally to his qubit:
\[
(I \otimes U_B)\ket{\chi_0}.
\]
Bob performs no measurement and stores no quantum information. He coherently modifies the shared entangled state and returns qubit $B$ to Alice through the same quantum channel.
This operation generalizes the original inter-family $(X,Y)$ Loop--Back mechanism to the full Pauli set, enabling both intra-family $(I,Z)$ and inter-family $(X,Y)$ transitions within the Bell basis.

The action of each Pauli operator on the Bell basis is summarized in Table~\ref{tab:pauli-bell-map} (global phases omitted), which defines the deterministic transition structure used for inference.

\begin{table}[h]
\centering
\caption{Action of local Pauli operations on Bell states. Global phases are omitted.}
\begin{tabular}{c|cccc}
\toprule
$U_B$ & $\Phi^{+}$ & $\Phi^{-}$ & $\Psi^{+}$ & $\Psi^{-}$ \\
\midrule
$I$ & $\Phi^{+}$ & $\Phi^{-}$ & $\Psi^{+}$ & $\Psi^{-}$ \\
$Z$ & $\Phi^{-}$ & $\Phi^{+}$ & $\Psi^{-}$ & $\Psi^{+}$ \\
$X$ & $\Psi^{+}$ & $\Psi^{-}$ & $\Phi^{+}$ & $\Phi^{-}$ \\
$Y$ & $\Psi^{-}$ & $\Psi^{+}$ & $\Phi^{-}$ & $\Phi^{+}$ \\
\bottomrule
\end{tabular}
\label{tab:pauli-bell-map}
\end{table}

\subsection*{Step 3: Bell-State Measurement at Alice}

After receiving the returned qubit, Alice performs a Bell-state measurement on the pair $(A,B)$.  
The Bell basis consists of the four maximally entangled states
\[
\begin{aligned}
\ket{\Phi^{\pm}} &= \tfrac{1}{\sqrt{2}}(\ket{00}\pm\ket{11}), &
\ket{\Psi^{\pm}} &= \tfrac{1}{\sqrt{2}}(\ket{01}\pm\ket{10}),
\end{aligned}
\]
implemented through the projective measurement operators
\[
\mathcal{M}_{\mathrm{Bell}}=\{P_{\Phi^{+}},P_{\Phi^{-}},P_{\Psi^{+}},P_{\Psi^{-}}\},
\qquad
P_{\chi}=\ket{\chi}\bra{\chi}.
\]
The measurement outcome is denoted by $\ket{\chi_{\mathrm{exp}}}$.

\subsection*{Step 4: Deterministic Inference of Bob's Operation}

Knowing the initial Bell state $\ket{\chi_0}$ and observing $\ket{\chi_{\mathrm{exp}}}$, Alice deterministically infers Bob's operation according to
\[
U_B = f(\chi_0,\chi_{\mathrm{exp}}),
\]
where $f$ is the inference rule induced by the Pauli transitions (Table~\ref{tab:pauli-bell-map}).  
For example, if $\chi_0=\ket{\Phi^{+}}$, the outcomes $\ket{\Psi^{+}}$ and $\ket{\Psi^{-}}$ correspond to the operations $X$ and $Y$, respectively.  
This inference capability exists solely because Alice retains knowledge of $\chi_0$, rendering the induced transition inaccessible to any adversary lacking the initial Bell reference.

\subsection*{Step 5: Determinism and Local Indistinguishability}

The protocol operates deterministically: each valid Bell-state measurement produces a usable inference event without basis matching or sifting.  
Since Bob's information is encoded in nonlocal entanglement correlations rather than in the traveling subsystem, any intermediate observer encounters a maximally mixed reduced state,
\[
\rho_B=\operatorname{Tr}_A(\ket{\chi_0}\bra{\chi_0})=\tfrac{I}{2},
\]
which carries no locally accessible information.  
Any interception, measurement, or substitution attempt necessarily perturbs the Bell-state statistics, producing anomalous distributions detectable during verification.

\subsection*{Step 6: Verification and Key Generation}

The above steps are repeated over $N$ independent rounds:
\begin{enumerate}
    \item In a randomly selected subset of rounds, Bob publicly reveals the applied operation $U_B$.
    \item Alice compares the announced values with her inferred results to estimate the quantum bit error rate (QBER).
    \item If the observed error rate remains within the expected experimental tolerance, the remaining rounds are retained for key generation through classical reconciliation and privacy amplification.
\end{enumerate}

\subsection*{Illustrative Example}

Consider the initial state
\[
\ket{\Phi^{+}} = \tfrac{1}{\sqrt{2}}(\ket{00} + \ket{11}),
\]
and the Pauli operators
\[
X = 
\begin{bmatrix}
0 & 1 \\ 
1 & 0
\end{bmatrix},
\qquad
Y = 
\begin{bmatrix}
0 & -i \\ 
i & 0
\end{bmatrix}.
\]

\paragraph{Case 1: Bob applies $X$.}
\[
(I \otimes X)\ket{\Phi^{+}} 
= \tfrac{1}{\sqrt{2}}(\ket{01} + \ket{10}) 
= \ket{\Psi^{+}}.
\]
Alice's Bell-state measurement yields $\ket{\Psi^{+}}$, identifying that Bob applied $X$.

\paragraph{Case 2: Bob applies $Y$.}
\[
(I \otimes Y)\ket{\Phi^{+}} 
= \tfrac{i}{\sqrt{2}}(\ket{10} - \ket{01}) 
= i\,\ket{\Psi^{-}}.
\]
The global phase $i$ is physically irrelevant, so the observable outcome is $\ket{\Psi^{-}}$, from which Alice infers that Bob applied $Y$.

The same inference principle applies to all Bell families.  
For instance, starting from $\ket{\Psi^{+}}$,
\[
(I \otimes X)\ket{\Psi^{+}} = \ket{\Phi^{+}}, 
\qquad
(I \otimes Y)\ket{\Psi^{+}} = i\,\ket{\Phi^{-}}.
\]
Thus, the inference rule remains consistent while the Bell family is inverted, confirming the inter-family nature of the process.

This behavior illustrates that the encoded information is carried by entanglement correlations rather than local quantum states.  
The protocol therefore provides deterministic inference symmetry across all Bell families and supports both key-distribution and direct-communication configurations, depending on the operational setting.

\section{Discussion: Algebraic Structure and Protocol Security}
\label{sec:discussion}

Let the Bell basis be defined as
\[
\begin{aligned}
\mathcal{B}_{\mathrm{Bell}} &=
\bigl\{\ket{\Phi^{\pm}},\,\ket{\Psi^{\pm}}\bigr\}, \\[4pt]
\ket{\Phi^{\pm}} &= \tfrac{1}{\sqrt{2}}(\ket{00}\pm\ket{11}),\quad
\ket{\Psi^{\pm}} = \tfrac{1}{\sqrt{2}}(\ket{01}\pm\ket{10}).
\end{aligned}
\]
The subsets $\{\Phi^{+},\Phi^{-}\}$ and $\{\Psi^{+},\Psi^{-}\}$ define two Bell
\emph{families}, distinguished by their parity structure.
The encoding properties of the proposed protocol arise from the way local
Pauli operations preserve or map between these families.
This algebraic structure underpins the operational security of the scheme
when the initial Bell state $\chi_0$ is selected privately by Alice.

\subsection*{Invariant Subspaces and Action of Pauli Operators}

The action of the Pauli operators on the second qubit spans the full
four-dimensional Bell space and induces both intra- and inter-family
transitions. Explicitly,
\[
\begin{aligned}
(I\!\otimes X)\ket{\Phi^{\pm}} &= \ket{\Psi^{\pm}}, &
(I\!\otimes X)\ket{\Psi^{\pm}} &= \ket{\Phi^{\pm}}, \\[3pt]
(I\!\otimes Z)\ket{\Phi^{\pm}} &= \ket{\Phi^{\mp}}, &
(I\!\otimes Z)\ket{\Psi^{\pm}} &= \ket{\Psi^{\mp}}, \\[3pt]
(I\!\otimes Y)\ket{\Phi^{\pm}} &= \pm i\,\ket{\Psi^{\mp}}, &
(I\!\otimes Y)\ket{\Psi^{\pm}} &= \pm i\,\ket{\Phi^{\mp}}.
\end{aligned}
\]
From these relations, two structural properties follow:
(P1) the operator $Z$ preserves each Bell family (\emph{intra-family}), and  
(P2) the operators $X$ and $Y$ induce transitions between Bell families
(\emph{inter-family}).

\subsection*{Intra-Family Encoding (\textit{Ping--Pong})}

In the \textit{Ping--Pong} protocol, Bob restricts his encoding to
$U_B\in\{I,Z\}$:
\[
(I\!\otimes I)\ket{\Psi^{+}}=\ket{\Psi^{+}},\qquad
(I\!\otimes Z)\ket{\Psi^{+}}=\ket{\Psi^{-}}.
\]
Information is encoded exclusively in the relative phase within a single
Bell family.
Once the initial state is publicly known, the effective signal space
$\mathcal{H}_{\Psi}=\mathrm{span}\{\ket{\Psi^{+}},\ket{\Psi^{-}}\}$ becomes
predictable.
This publicly accessible symmetry allows certain coherent or
family-preserving substitution strategies to remain confined within the
same two-dimensional subspace, delaying their detectability.

\subsection*{Inter-Family Encoding (\textit{Loop--Back Bell-State})}

In the proposed \textit{Loop--Back Bell-State} protocol, Bob applies
$U_B\in\{X,Y\}$ to an arbitrary initial state
$\chi_0\in\mathcal{B}_{\mathrm{Bell}}$:
\[
(I\!\otimes X)\ket{\Phi^{+}}=\ket{\Psi^{+}},\qquad
(I\!\otimes Y)\ket{\Phi^{+}}=i\,\ket{\Psi^{-}},\quad\text{etc.}
\]
When $\chi_0$ is chosen uniformly at random and kept secret by Alice, the
mapping $(\chi_0,U_B)\mapsto\chi_{\mathrm{exp}}$ becomes inaccessible to any
external observer.
Although the reduced traveling state
\[
\rho_B=\operatorname{Tr}_A(\ket{\chi_0}\bra{\chi_0})=\tfrac{I}{2}
\]
ensures local indistinguishability, the essential security feature lies in
the absence of a public reference frame required to interpret Bell-family
transitions.
Without knowledge of $\chi_0$, no adversary can consistently reproduce
Bob's operation across rounds without introducing detectable statistical
inconsistencies.

The distinction between intra- and inter-family transformations plays a central role in the present protocol. Similar Bell-family separations have previously been exploited in the context of intrinsic verification of Bell-basis quantum computation, where inter-family transitions act as witnesses of global consistency~\cite{lizama2025bell}. Here, the same structural separation is leveraged in a cryptographic setting to amplify detectability and remove public reference frames.

Table~\ref{tab:comparison_protocols} compares the proposed protocol with representative two--way and Loop--Back QKD schemes, highlighting differences in encoding space, reference-state disclosure, and remote-node complexity.

\begin{table}[t]
\centering
\caption{Comparison between representative two--way and Loop--Back QKD protocols.}
\label{tab:comparison_protocols}

\setlength{\tabcolsep}{4pt}
\renewcommand{\arraystretch}{1.15}

\begin{tabular*}{\textwidth}{@{\extracolsep{\fill}}
P{0.24\textwidth} P{0.16\textwidth} P{0.16\textwidth} P{0.18\textwidth} P{0.18\textwidth}}
\toprule
\textbf{Feature} & \textbf{PP} & \textbf{LM05} & \textbf{LB--3B} & \textbf{This work} \\
\midrule
Quantum resource & Bell states & 1 qubit & 1 qubit & Bell states \\
Encoding space dimension & $2$ & $2$ & $2$ & $4$ \\
Reference state & Public Bell family & Public state & Public qubit state & \textbf{Private Bell state} \\
Bob's role & Active & Active & Passive & \textbf{Passive} \\
Bidirectional propagation & Yes & Yes & Yes & Yes \\
Detectors at Bob & Yes & Yes & No & \textbf{No} \\
\bottomrule
\end{tabular*}

\vspace{0.7em}

\begin{tabular*}{\textwidth}{@{\extracolsep{\fill}}
P{0.24\textwidth} P{0.16\textwidth} P{0.16\textwidth} P{0.18\textwidth} P{0.18\textwidth}}
\toprule
\textbf{Feature} & \textbf{PP} & \textbf{LM05} & \textbf{LB--3B} & \textbf{This work} \\
\midrule
Basis sifting & No & No & No & No \\
Intrinsic efficiency & $50\%$ & $50\%$ & $\sim 25$--$26\%$ & \textbf{$\leq p_{\mathrm{BSM}}$} \\
Substitution detectability & Limited & Limited (IR) & Statistical & \textbf{Amplified ($P_d\simeq 3/4$)} \\
Mobile / edge suitability & Low & Moderate & High & \textbf{High} \\
\bottomrule
\end{tabular*}

\vspace{0.3em}
\footnotesize\textit{PP: Ping--Pong; LB--3B: Loop--Back three--basis; IR: intercept--resend.}

\end{table}

\subsection*{Group Action and Hidden Reference Structure}

The algebraic structure underlying the proposed protocol can be formalized as a group action on the Bell-state space.
Let $\mathcal{P}=\{I,X,Y,Z\}$ denote the single-qubit Pauli group modulo global phases, and let
$\mathcal{B}_{\mathrm{Bell}}=\{\Phi^{+},\Phi^{-},\Psi^{+},\Psi^{-}\}$ be the Bell basis.
The local action of $\mathcal{P}$ on the second qubit induces a transitive group action
\[
\alpha:\mathcal{P}\times\mathcal{B}_{\mathrm{Bell}}\longrightarrow\mathcal{B}_{\mathrm{Bell}},
\qquad
\alpha(U,\chi)=(I\otimes U)\ket{\chi},
\]
which maps any Bell state to another element of the same basis.
As shown in Table~\ref{tab:pauli-bell-map}, this action spans the full four-dimensional Bell space and generates both intra-family and inter-family transitions.

From Alice's perspective, the protocol defines an invertible mapping
\[
(\chi_0,U_B)\longmapsto \chi_{\mathrm{exp}},
\]
where $\chi_0$ is the initial Bell state privately chosen by Alice, $U_B\in\mathcal{P}$ is Bob's local operation, and $\chi_{\mathrm{exp}}$ is the Bell-state measurement outcome.
Since Alice knows $\chi_0$, she can deterministically recover $U_B$ by inverting the action $\alpha$.

In contrast, for any external observer lacking access to $\chi_0$, the Bell state serves as a hidden reference frame.
If $\chi_0$ is selected uniformly at random from $\mathcal{B}_{\mathrm{Bell}}$, then for any fixed $U_B$ the conditional distribution of the observed outcome satisfies
\[
\Pr(\chi_{\mathrm{exp}}\,|\,U_B)=\tfrac{1}{4}, \qquad \forall\,\chi_{\mathrm{exp}}\in\mathcal{B}_{\mathrm{Bell}}.
\]
Thus, without knowledge of the initial reference state, the action of $U_B$ is statistically indistinguishable, even though the local reduced state $\rho_B=I/2$ remains maximally mixed for all protocol rounds.

This hidden-reference structure is a central security feature of the protocol.
While the Pauli group action is fully deterministic for the legitimate parties, it is rendered non-invertible for an adversary who does not possess the reference $\chi_0$.
As a consequence, any attempt to replicate Bob's operation or to perform a family-preserving substitution necessarily breaks the consistency of the induced group action across rounds, leading to statistically detectable anomalies in the Bell-state distribution.

\subsection*{Security Implications}

A key distinction with respect to established two-way protocols concerns
the detectability of separable substitution attacks.
In subspace-restricted schemes such as \textit{Ping--Pong}, a carefully
chosen substitution that preserves the Bell family may remain confined
within a publicly known two-dimensional space.
In contrast, the proposed protocol forces the honest evolution to explore
the entire four-dimensional Bell basis.
As a result, any separable substitution $\sigma_B$ yields an output
distribution that approaches uniformity over
$\mathcal{B}_{\mathrm{Bell}}$, leading to a per-round detection probability
on the order of $P_d\simeq 3/4$.

\medskip\noindent\textbf{Operational Indistinguishability.}\par
Because $\rho_B=I/2$ for all $\chi_0$ and $U_B$, local measurements on the
traveling qubit cannot reveal the applied operation without disturbing the
entanglement.
Intercept--resend or entanglement-swapping attacks therefore reduce the
fidelity
\[
F=\bra{\chi_{\mathrm{exp}}}\rho_{AB}\ket{\chi_{\mathrm{exp}}},\qquad
Q=1-F,
\]
manifesting as an increased quantum bit error rate (QBER).

\medskip\noindent\textbf{Full-Pauli Encoding.}\par
Extending the encoding to the full Pauli set does not alter local
indistinguishability.
The secrecy of $\chi_0$ prevents exploitation of intra-family symmetries,
ensuring that both intra-family $(I,Z)$ and inter-family $(X,Y)$
transitions become indistinguishable to an adversary lacking access to
Alice's qubit.

\medskip\noindent\textbf{Signal Space and Detectability.}\par
While \textit{Ping--Pong} operates within an invariant two-dimensional
subspace, the present protocol spans the full Bell space.
Consequently, separable substitution attacks induce statistically
inconsistent Bell-state distributions, resulting in a detection
probability approaching $P_d\simeq 3/4$ per round.

\medskip\noindent\textbf{Untrusted Source.}\par
The protocol does not require trust in the entanglement source.
Alice and Bob may verify nonlocal correlations by measuring in complementary bases and checking for either a violation of a Bell inequality, such as the CHSH criterion~\cite{clauser1969proposed}, or a fidelity threshold $F>F_{\min}$.
This approach follows the entanglement-based QKD paradigm originally proposed by Ekert~\cite{ekert1991quantum}.

\section{Future Work and Perspectives}

The practical implementation of the \textit{Loop--Back Bell-State QKD} protocol relies on the availability of high-fidelity entanglement sources and reliable Bell-state measurements.
Current photonic platforms already provide most of the required ingredients, including integrated sources of entangled photon pairs, reconfigurable linear-optical circuits, and Bell-state analyzers operating with success probabilities close to $p_{\mathrm{BSM}}\simeq 0.5$~\cite{wang2020integrated}.
Recent demonstrations using photon-number-resolving detectors and highly indistinguishable single-photon emitters have shown that near-deterministic Bell measurements are becoming experimentally accessible~\cite{bayerbach2023bell,grice2011arbitrarily,ewert20143}.
These advances suggest that the proposed protocol is compatible with near-term photonic implementations.
In practical realizations based on spontaneous parametric down-conversion (SPDC) sources, the scheme may be combined with decoy-state techniques or photon-number monitoring to mitigate multi-photon vulnerabilities without altering the core loop--back architecture~\cite{hwang2003quantum,lo2005decoy}.

Beyond photonic systems, alternative platforms such as trapped ions, superconducting qubits, and atom-based quantum memories offer nearly deterministic local unitaries and high-fidelity joint measurements.
These features make them natural candidates for exploring the loop--back mechanism in modular or hybrid quantum-network settings, where the inter-family encoding introduced here could support secure synchronization links or distributed key establishment among multiple nodes.

From a theoretical perspective, several directions merit further investigation.
A first avenue concerns the analysis of the protocol under asymmetric or correlated noise models, where inter-family and intra-family transitions may exhibit distinct robustness properties.
A second direction involves the study of generalized local encoding operations $U\in SU(2)$, with the aim of identifying which classes of unitaries maximize statistical indistinguishability in the Bell space and potentially improve error tolerance.
Finally, the deterministic Bell-state transitions underlying the protocol could be incorporated into quantum secure direct communication or key-recycling frameworks, where the verification of nonlocal correlations may provide additional layers of authentication or message-integrity checking.

A full secret-key-rate (SKR) derivation based on the Devetak--Winter bound~\cite{devetak2005distillation} is beyond the scope of the present work, which focuses on the architectural and algebraic properties of the protocol.
Nevertheless, at a heuristic level, the SKR can be expressed in the standard form
\[
R \geq p_{\mathrm{BSM}} \, \bigl[ 1 - h(Q) - I_{\mathrm{Eve}} \bigr].
\]
where $p_{\mathrm{BSM}}$ denotes the Bell-state measurement success probability, $Q$ is the observed quantum bit error rate, and $I_{\mathrm{Eve}}$ represents an upper bound on Eve's accessible information.
In the context of separable substitution or intercept--resend strategies, $I_{\mathrm{Eve}}$ is constrained by the local indistinguishability of Pauli operations under $\rho_B = I/2$ and by the absence of a public reference Bell state.
Under these conditions, the protocol exhibits an enhanced disturbance sensitivity, with a characteristic detection probability on the order of $3/4$.
A rigorous optimization of $R$ under realistic noise models and general attack strategies is left for future work.

\section{Conclusions}

We have presented a Bell--state extension of the Loop--Back quantum key distribution architecture, preserving its defining features---bidirectional propagation, a simplified and measurement--free remote terminal, and intrinsic channel verification---while expanding the effective signal space through entanglement correlations.
By moving from single--qubit encodings to Bell--state transitions, the proposed scheme lifts a structural limitation of earlier Loop--Back realizations, which were restricted to two--dimensional Hilbert spaces and exhibited an intrinsic post--selection efficiency of approximately $25$--$26\%$.

The protocol employs deterministic local Pauli operations at the remote node, inducing well--defined transitions within the Bell basis that are unambiguously resolved by a Bell--state measurement performed at Alice's station.
A central architectural element is that Alice privately prepares and retains knowledge of the initial Bell state, which serves as a hidden reference for interpreting the observed transition.
An adversary lacking access to this reference cannot reconstruct the mapping between the applied operation and the measured Bell state, even though the traveling subsystem remains locally maximally mixed, $\rho_B = I/2$, throughout the protocol.

This structural asymmetry reinforces operational indistinguishability and provides a built--in mechanism for detecting non--honest evolutions.
Any substitution or replication attempt that does not preserve the correct Bell--state transition pattern necessarily produces statistically inconsistent outcomes during verification.
In this sense, the protocol complements the intrinsic intercept--resend detectability of earlier Loop--Back schemes with an additional layer of robustness arising from the enlarged entangled--state signal space.

From an efficiency perspective, the proposed extension preserves the operational simplicity of Loop--Back QKD while removing its single--qubit post--selection bottleneck.
The effective throughput is no longer constrained by basis reconciliation, but instead bounded by the success probability of the Bell--state measurement, $p_{\mathrm{BSM}}$.
In standard linear--optical implementations, this corresponds to efficiencies of up to $50\%$, with the potential to approach deterministic operation in platforms supporting near--unit Bell--state discrimination.

Overall, the Loop--Back Bell--state protocol establishes a coherent architectural pathway for enhancing efficiency and robustness in two--way QKD without sacrificing passive operation at the network edge.
By lifting the intrinsic efficiency ceiling of single--qubit Loop--Back schemes, the proposed Bell--state extension removes a key practical obstacle for mobile QKD, where passive remote nodes must operate under high loss and limited temporal availability.
In this way, the scheme provides a flexible foundation for future multi--node, mobile, or hybrid quantum--network scenarios in which minimal classical disclosure and asymmetric hardware requirements are operationally advantageous.

\end{document}